 \documentclass[final,5p,times,twocolumn,authoryear]{elsarticle}

\usepackage{amssymb}
\usepackage{lipsum}

\usepackage{natbib}
\usepackage{overpic}

\usepackage{color,soul}
\usepackage{booktabs,xcolor}

\journal{Materials Today Communications}

\begin{document}

\begin{frontmatter}



\title{Cerium doped graphene-based materials towards oxygen reduction reaction catalysis}


\author{Lanna E. B. Lucchetti\fnref{label1}}
\affiliation[label1]{organization={Centro de Ciências Naturais e Humanas (CCNH) - Universidade Federal do ABC (UFABC)},
            city={Santo André},
            state={SP},
            country={Brazil}}
            
\author{Pedro A. S. Autreto\fnref{label1}}
\author{Mauro C. Santos\fnref{label1}}
\author{James M. de Almeida\fnref{label2}}
\affiliation[label2]{organization={Ilum School of Science (CNPEM)},
            city={Campinas},
            state={SP},
            country={Brazil}}

\begin{abstract}
With the global transition towards cleaner energy and sustainable processes, the demand for efficient catalysts, especially for the oxygen reduction reaction, has gained attention from the scientific community. This research work investigates cerium-doped graphene-based materials as catalysts for this process with density functional theory calculations. The electrochemical performance of Ce-doped graphene was assessed within the computation hydrogen electrode framework. Our findings reveal that Ce doping, especially when synergized with an oxygen atom, shows improved catalytic activity and selectivity. For instance, Ce doping in combination with an oxygen atom, located near a border, can be selective for the 2-electron pathway. Overall, the combination of Ce doping with structural defects and oxygenated functions lowers the reaction free energies for the oxygen reduction compared to pure graphene, and consequently, might improve the catalytic activity. This research sheds light from a computational perspective on Ce-doped carbon materials as a sustainable alternative to traditional costly metal-based catalysts, offering promising prospects for green energy technologies and electrochemical applications.
\end{abstract}




\end{frontmatter}




\section{Introduction}
\label{introduction}

As the world shifts toward cleaner and more sustainable energy generation, the development of efficient catalysts is critical \citep{sustainable}. Electrochemical processes play a pivotal role in a wide range of energy conversion technologies and green chemistry applications, such as fuel cells \citep{fuelcell} and water treatment \citep{fenton}. 
The oxygen reduction reaction (ORR) is particularly important in that context \citep{energy}. When this reaction follows a 4-electron mechanism, oxygen molecules are electrochemically reduced to water, releasing energy that can be harnessed to generate electricity. On the other hand, hydrogen peroxide is obtained when it follows a 2-electron mechanism, a very attractive alternative to the harmful anthraquinone process \citep{h2o2}. Regardless, the ORR is central to the operation of many clean energy systems and greener processes, and its efficiency is of paramount importance for the practical viability of these technologies. Traditional catalysts used in these processes, however, are usually based on rare and expensive metals. Therefore, new catalysts based on more abundant and sustainable materials are detrimental to reducing the environmental impact of the new processes and enhancing their sustainability. Furthermore, the novel catalysts also have to promote a good electrochemical performance of ORR both in terms of activity and selectivity. 

In this sense, recent progress in materials science and computational modelling techniques has opened up new possibilities for designing catalysts with precisely tailored structures and properties. This design-driven approach allows researchers to optimize catalysts for specific applications, potentially achieving breakthroughs in ORR efficiency. Additionally, metal-doped graphene systems have been exhaustively investigated with theoretical calculations \citep{LUCCHETTI2021115429, mcatalysts, mcatalysts2}. However, because most studies focus on transition metal doping, Ce-doped carbon surfaces remain to be explored. Some studies in that direction have investigated different catalytic reactions on Ce single atoms adsorbed on carbon nanotubes \citep{nanotube}, Ce atoms adsorbed on graphene flakes \citep{flake}, and Ce-N-coordinated graphene \citep{CeNgraph}. Other works combine cerium with carbon nitride \citep{cn, cn2} or boron nitride sheets \citep{bn}. Besides aiming to address a less explored route in metal functionalization for graphene, this work is also motivated but a contemporary issue regarding rare earth metals mining. Ce and La make up for 70$\%$ of the mined rare earth metals, and Ce is the predominant element in rare earth ores \citep{cerium, binnemans2015rare}. However, despite its availability, Ce is still treated as an undesirable and a low-value byproduct. Therefore, investigating and developing novel applications to employ Ce might eventually induce a positive effect on the waste and environmental impacts of the mining industry, while simultaneously providing cheaper and relatively abundant catalysts compared to precious or scarce transition metals.

Herein, the catalytic activity of different graphene-based materials doped with Ce in combination with structural defects, oxygenated functions, and borders, is assessed with density functional theory (DFT) calculations. Our results show that Ce doping can improve the catalytic activity of the carbon sheets. Particularly, the combination of cerium doping with an oxygen atom has the best performance, indicated by the lowest theoretical overpotential values for both 4-electron and 2-electron ORR. When this functionalization is close to a border, the 2-electron ORR overpotential becomes significantly lower, indicating a promising site to promote H$_2$O$_2$ electrogeneration selectively. This investigation has been carried out to fill in a gap in the theoretical literature and to shed light on the role of Ce doping in the catalytic activity of graphene towards the ORR. 


\section{Methods}
\label{methods}
\begin{figure}
	\centering 
	\includegraphics[width=0.5\textwidth, angle=0]{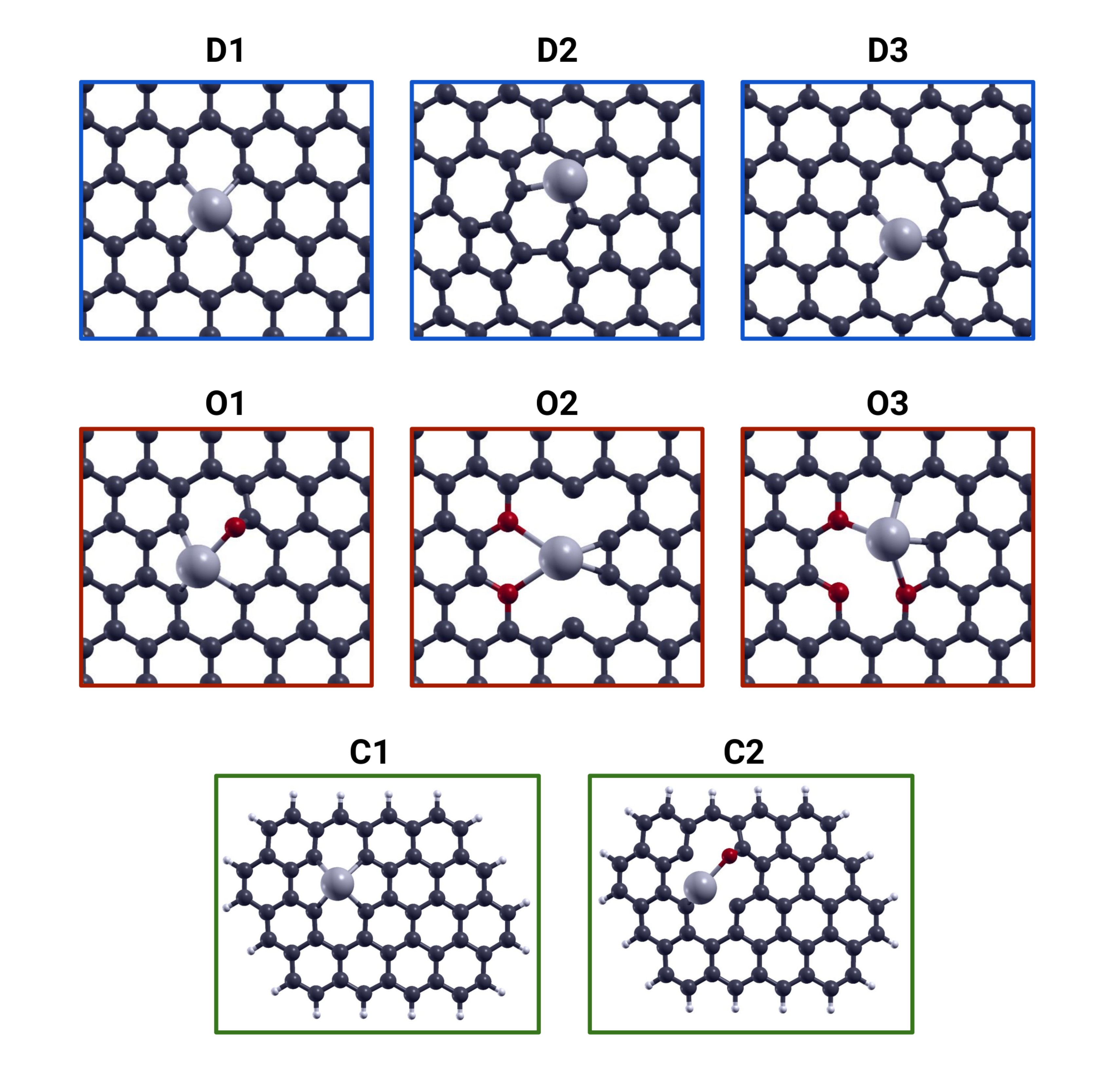}	
	\caption{Selected structural models of Ce doping combined with different defects (blue framed), oxygenated functions (red framed), and clusters (green framed). The defects are, respectively: a carbon double vacancy (D1), a 555-6-777 defect (D2), and a 555-777 defect (D3); Ce coordinated with one (O1), two (O2) and three (O3) oxygen atoms; a carbon double vacancy (C1) and Ce coordinated with one oxygen atom (C2) clusters. } 
	\label{fig1}%
\end{figure}

We have investigated different combinations of Ce functionalization with 1) defects on the carbon network — double vacancy, 555-6-777 and 555-777; 2) oxygenated functions — Ce coordinated with one, two, or three oxygen atoms; and 3) finite clusters containing either a double-vacancy or its combination with one oxygen atom. Atomically dispersed cerium atoms on graphene have been previously reported and investigated by \citep{cerium-graphene-2018} and by \citep{flake}. Furthermore, 5-9 nm wide CeO$_2$ nanoparticles, which is very close to a single Ce coordinated with O atoms on graphene, have been synthesized by \citep{Cu-Ce-nano}, combined with Cu and graphene with great catalyst stability being reported by the authors. Furthermore, cerium incorporated to graphene was obtained by \cite{LI2022128048}. These experimental works illustrate the stability of similar structures, even though the exact nature of the catalytically active Ce site coordination remains to be determined. The different functionalizations proposed in our work have been based on experimental findings, like the aforementioned ones and by similar computational investigations carried out for different single metal doping graphene, such as single atom Fe-doped graphene \citep{norskov-tm-sac}. We reckon that there are numerous possibilities of vacancies, defects and oxygenated functions combined with single-atom catalysts, making it impossible to address all at once. We approached this issue by selecting different combinations that have not yet been reported in the literature for Ce, but have shown good catalytic activity with other metals \citep{LUCCHETTI2021115429, norskov-tm-sac}. Figure \ref{fig1} illustrates the structural models considered in this work and these structures will be referred to as D1, D2, D3 (defects); O1, O2, O3 (oxygenated functions), C1 and C2 (clusters). 

To optimize the selected structures and access their catalytic activity towards the ORR, plane-wave DFT calculations were performed with the Quantum ESPRESSO package \citep{Giannozzi_2009}. Standard solid-state efficiency pseudopotentials (SSSP) were employed to describe core electrons \citep{sssp}, and plane-wave basis functions with a 50 Ry kinetic energy cutoff to describe valence electrons. The exchange-correlation energy was treated with generalized gradient approximation functionals (GGA-PBE) \citep{perdew1996generalized}. A Hubbard-like parameter of $U_{eff} = 5 eV$ was applied to correct the on-site Coulomb interaction of localized Ce-4$f$ states \citep{hub}. The ORR was modelled to address the two possible reaction mechanisms, according to the following equations \citep{norskov2004}:

\begin{equation}
\centering
    O_2 + 4\hspace{2pt}(H^+ + e^-) \rightarrow 2 \hspace{2pt}H_2O
    \label{eq1}
\end{equation}

\begin{equation}
    O_2 + 2\hspace{2pt}(H^+ + e^-) \rightarrow H_2O_2
    \label{eq2}
\end{equation}
\vspace{1.5pt}

The reaction mechanism is given by:

\begin{equation}
    O_2 + (H^+ + e^-) + * \rightarrow *OOH
\end{equation}

\begin{equation}
    *OOH + (H^+ + e^-) \rightarrow H_2O  + *O
\end{equation}

\begin{equation}
    *O + (H^+ + e^-) \rightarrow *OH
\end{equation}

\begin{equation}
    *OH + (H^+ + e^-) \rightarrow H_2O
\end{equation}


where (*) indicates a surface site. The *OOH is a key intermediate after which a  H$_2$O$_2$ molecule can be formed on the 2-electron pathway, or it can undergo further reduction along with the O-O bond break, following the 4-electron pathway. This reaction can be modeled within the computational hydrogen electrode (CHE) framework proposed by Nørskov \textit{et al.} \citep{norskov2004}. In this case, the intermediate reaction steps, namely *OOH, *O, and *OH are simulated with DFT adsorption calculations. The adsorption energies are calculated taking gaseous H$_2$ and H$_2$O molecules as references, considering the system to be in equilibrium and at $pH = 0$. Under these assumptions, the energy of a proton/electron pair can be given by:

\begin{equation}
    (H^+ + e^-) = \frac{1}{2} H_2
\end{equation}

The experimental reduction potential values are also considered, to determine the total reaction Gibbs free energy change - $E^0 = 1.23 V$, for the 4-electron and $E^0 = 0.70 V$ for the 2-electron ORR. Finally, the theoretical overpotential ($\eta$), used to describe the catalytic activity is calculated according to the following equation :

\begin{equation}
  \eta = \frac{max[\Delta G_1, \Delta G_2, \Delta G_3, \Delta G_4]}{e}
  \label{n}
\end{equation}

The limiting potential ($U_{lim}$) is derived from Equation \ref{n} as follows:

\begin{equation}
  U_{lim} = E^0 - \eta
\end{equation}

\section{Results and Discussion}

\begin{figure*}
	\centering 
	\includegraphics[width=0.9\textwidth, angle=0]{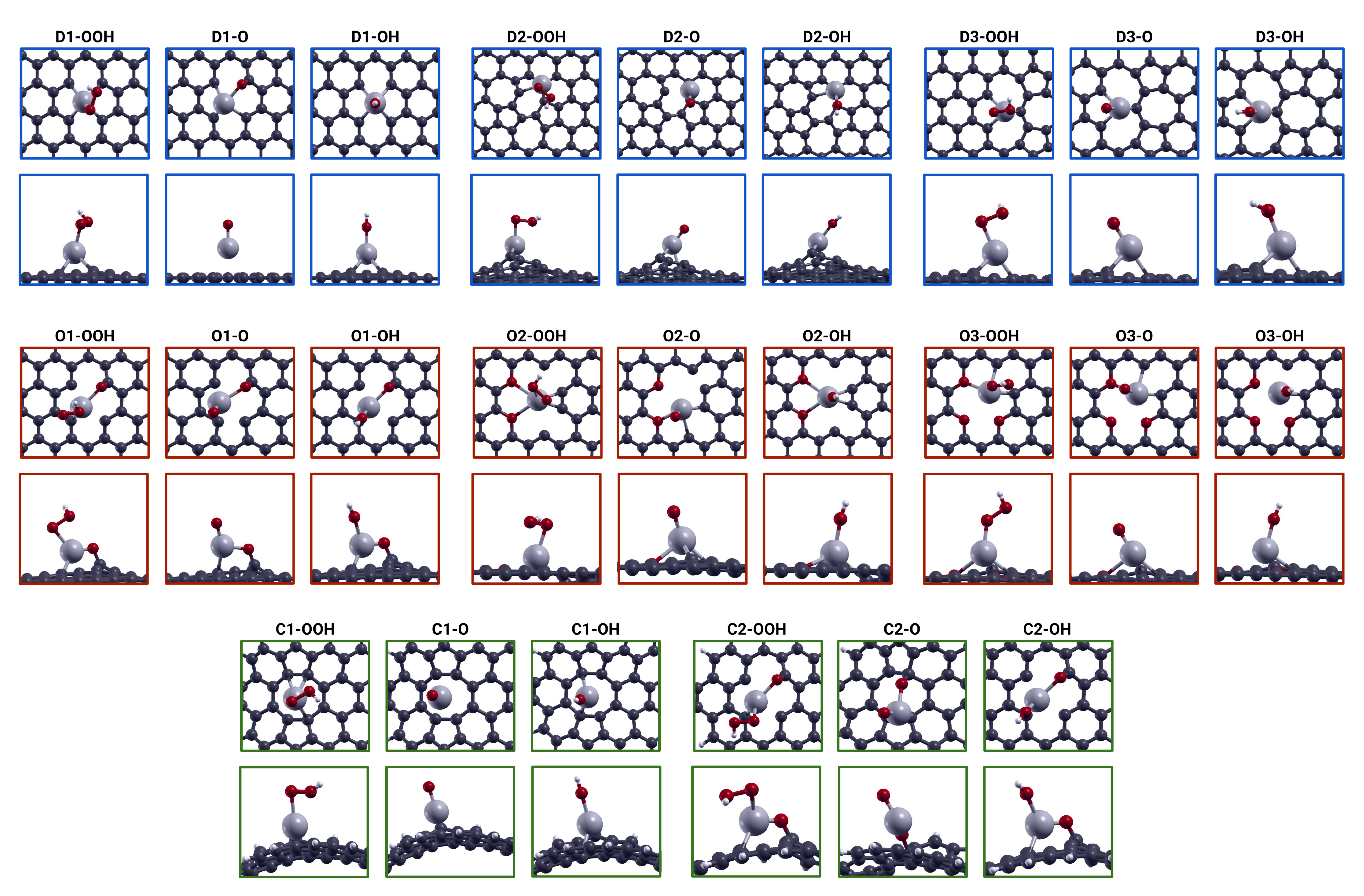}	
	\caption{Optimized coordinates of oxygen reduction reaction steps on different Ce-functionalized graphene.} 
	\label{fig2}%
\end{figure*}

The stability of the proposed functionalizations in this work has been investigated in terms of formation energies. The formation energy was determined by taking the obtained surface and subtracting the total energy of its isolated compounds - the graphene sheet, without the functionalization, and the O and Ce atoms, taking the O$_2$ molecule, when applicable, and the cubic Ce Fm3m structure as a reference \cite{mp}. This is summarized by the following equation:

\begin{equation}
    E_{form} = E_{structure} - E_{graphene} - E_{Ce}
\end{equation}

Employing this equation, negative formation energy values will indicate that the obtained structure is energetically favourable in comparison with its isolated compounds. Table 1 shows the obtained formation energies and it can be seen that the adsorption of the Ce functionalization to the defected graphene sheets is favourable in all cases proposed in this work.

\begin{table}[!h]
\label{tab1}
\begin{tabular}{l c c c} 
 \hline
 Surface & $E_{form} (eV)$   \\
 \hline
D1 & -4.45  \\
D2 &  -3.72 \\
D3 &  -4.80  \\
O1 &  -6.20   \\
O2 &   -10.13  \\
O3 &  -10.82   \\
C1 &  -4.94  \\
C2 &  -7.67  \\ 
 \hline
 \end{tabular}
 \caption{Calculated formation energies of the different functionalizations proposed in this work. }

\end{table}

\begin{figure}
\centering 
	\includegraphics[width=0.49\textwidth, angle=0]{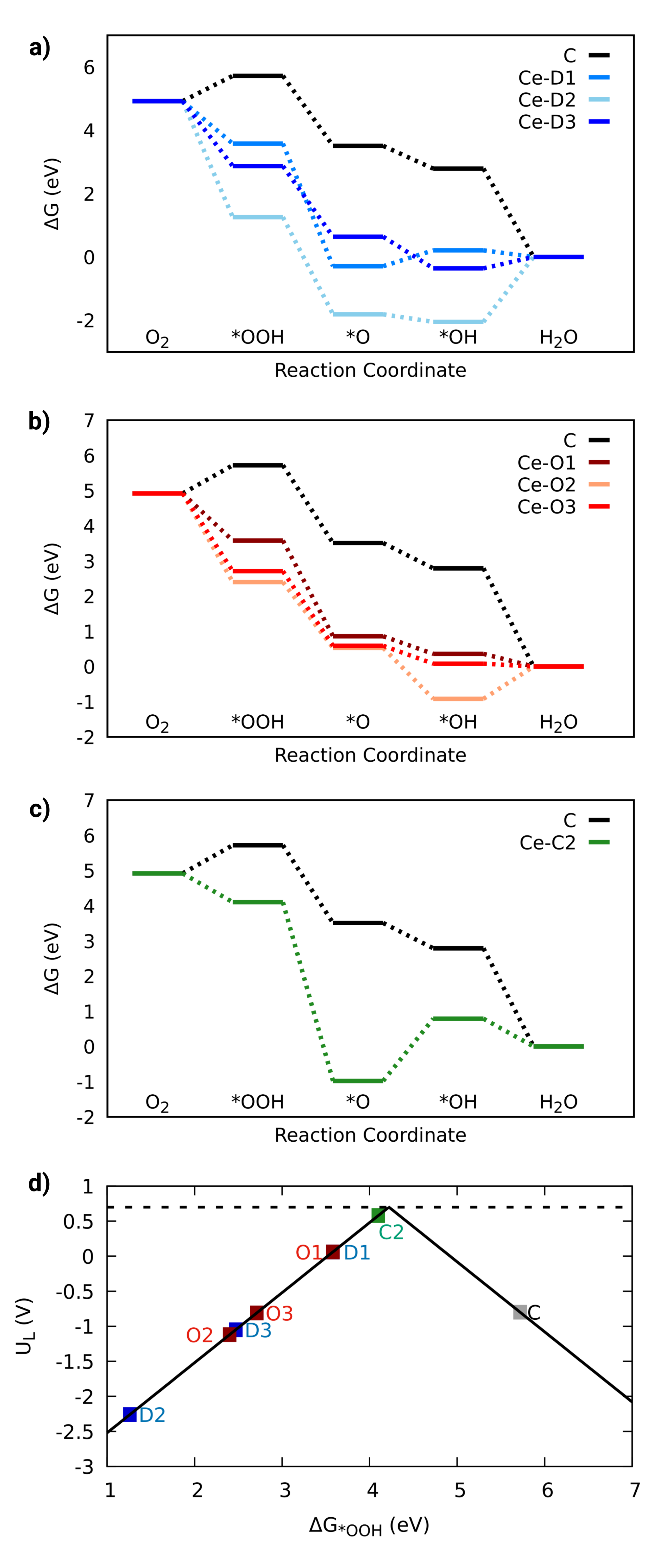}
	\caption{Gibbs free energy diagrams on different Ce-functionalized graphene (a-c) and volcano plot of the catalytic activity in terms of *OOH Gibbs free energy and limiting potential for the 2-electron pathway (d). } 
	\label{fig3}%
\end{figure}

Figure \ref{fig2} shows top and side views of all ORR intermediate reaction steps (*OOH, *O, and *OH) on the different structures after relaxation. The Gibbs free energy diagrams in Figure \ref{fig3} and the calculated theoretical overpotential values in Table \ref{tab2} show that Ce-doping lowers the adsorption energy of all ORR intermediates compared to pristine graphene. When Ce is combined with a double vacancy or a 555-777 defect (D1 and D3), as shown in Figure \ref{fig3}a, the *OOH adsorption free energy values become closer to the ideal. However, the interaction with the *O intermediate is too strong and that could lead to surface poisoning or hinder the catalytic activity altogether. On D2, the adsorption energy of all intermediate species is below the acceptable range for a catalytic material. This can be attributed to the considerable surface distortion that happens along with the reaction steps in this case. Interestingly, D1 has a theoretical overpotential of $\eta = 0.64 \hspace{2pt}V$ for the 2-electron pathway and $\eta = 1.73 \hspace{2pt}V$ for the 4-electron pathway. While $\Delta G_{*OOH}$ is ideal on this surface, $\Delta G_{*O}$ is too negative, making the O-O bond break the potential determining step. The strong interaction between the Ce and the O atom also results in a bond stretching to the point that the Ce atom moves away from the carbon network, indicating that further stabilizing would be required in this case. This could be achieved by surrounding the Ce atom with more electronegative species, such as oxygen itself, or nitrogen. The overpotential value $\eta = 0.64 \hspace{2pt}V$ is still higher than ideal, but it indicates that this structure could be selective for the 2-electron ORR.

\begin{table}
\label{tab2}%
\begin{tabular}{l c c c} 
 \hline
 Surface & $\eta(4e^-)$  & $\eta(2e^-)$   \\
 \hline
pristine & 2.03 &  1.50  \\
D1 & 1.73 & 0.64   \\
D2 & 3.28 & 2.96   \\
D3 & 1.94 & 1.75   \\
O1 & 0.87 & 0.64   \\
O2 & 2.15 & 1.82   \\
O3 & 1.15 & 1.51   \\
C1* & n/a &  n/a   \\
C2 & 3.00 & 0.12   \\ 
 \hline
 \end{tabular}
 \caption{Calculated theoretical overpotentials for the 4-electron and 2-electron oxygen reduction reaction on different Ce-functionalized graphene. \\ (*)The ORR species are not stable and desorb from the structure forming a defect in this case.}

\end{table}

On the other hand, when Ce is combined with oxygenated functions, the catalytic activity can be tuned accordingly. The O1 structure particularly approaches the ideal catalytic behavior (Figure \ref{fig3}b), lowering the reaction free energy values. It can also be noted that the oxygen coordination indeed promotes additional stabilization of the Ce atom, which does not move away from the carbon support in this case (Figure \ref{fig2}). The 2-electron overpotential is similar to the D1 structure, and the 4-electron ORR has an overpotential of $\eta = 0.87 \hspace{2pt}V$. As the number of oxygen atoms increases, however, the interaction between the Ce atom and the ORR intermediates becomes too strong, which worsens the catalytic performance. Among the three structures considered in this case, O1 is the most promising for ORR applications via either 2-electron or 4-electron pathways.

The catalytic profile of Ce-doped graphene combined with either structural defects or oxygen atoms is overall very similar to what has been observed by Liu \textit{et al.} \citep{CeNgraph}. The authors noted that cerium-doped graphene coordinated by nitrogen atoms had better catalytic activity than pristine graphene, with negative Gibbs free energy following the reaction steps. However, the calculated theoretical overpotential values were still too high for most Ce-N-doped graphene structures due to the considerably strong absorption of ORR intermediate species. The only exception among the investigated structures in this work was Ce(OH)-N$_4$-C, where the Ce atom was also coordinated with an -OH axial ligand, with comparable catalytic performance to Fe-N$_4$ doped graphene. In fact, other works in the literature suggest that not only the -OH axial ligand can improve the catalytic activity, but it can also shift the preferred pathway \citep{feNOH, ruNOH}, which can be a promising direction for future studies of Ce-doped carbon materials. 

On the other hand, when the ORR intermediates are adsorbed on the double-vacancy Ce-doped cluster (C1) the oxygen atom forms a strong bond with the cerium atom and it desorbed from the surface. The strong bonds yield very negative adsorption free energy values, and as the species desorbed, Stone-Wales defects are formed in all cases (Figure \ref{fig2}). This means the ORR is not favorable and this would be a poor catalytic site for this reaction. However, it can be seen that when a Ce atom is close to a border and interacts with oxygenated species forming a Stone-Wales defect, the carbon network reactivity can change. This is aligned with experimental works in the literature, where carbon-based catalysts had more defects and oxygenated groups after their modification with ceria nanostructures \citep{cec, cec1}. 

Furthermore, \citep{flake} have shown that Ce/Cd nanocatalysts exhibited excellent performance and stability towards the 4-nitrophenol reduction and olefin oxidation reactions. Additionally, a work from \citep{Cu-Ce-nano} presented a synthesized Cu-Ce graphene catalysts for CO oxidation where the experimental characterization via XRD, XPS, and Raman spectra indicated that the well dispersed Ce could be found in combination with oxygenated species. The authors state that the interaction with dispersed Ce and graphene is very significant, stronger than that of Cu, providing experimental evidence of the stability of these structures. Their results indicate that the catalysts prepared in this work would perform poorly for the CO oxidation reaction when those were single metal-supported catalysts, namely Cu/graphene and Ce/graphene, whereas the bimetallic catalysts had a better performance, attributed to the relative number of active species and oxygenated groups on the catalytic surface, which were determined by the Cu/Ce molar ratio. These experimental findings can be correlated with what has been observed in this work, since our DFT calculations show that different proposed sites yielded very different catalytic profiles, and the combination of Ce with a second metallic atom might have induced changes in the catalyst structure and its specific sites, therefore shifting the catalytic activity, as the authors indicate.

Similar to what was observed for the periodic system, a neighboring oxygen added to the C2 model can stabilize the Ce atom and the cluster retains its structure when the *OOH intermediate is adsorbed on this site. A high theoretical overpotential for the 4-electron pathway is observed, nonetheless, due to a strong interaction with the *O intermediate species and a different carbon defect formed in this case. The $\Delta G_{*OOH}$ value, however, is ideal for the 2-electron pathway, yielding a very low overpotential value of  $\eta = 0.12 \hspace{2pt}V$. If *OOH is taken as a descriptor, a volcano plot can be built, so the catalytic activity can be evaluated in terms of the limiting potential, as shown in Figure \ref{fig3}d. Pristine graphene, indicated by the gray data point, is also plotted for comparison. It can be seen that most of the selected structures have poor catalytic activity with unfeasible limiting potentials, like D2, D3, O2, and O3. Three configurations are promising – D1 and O1, where the reaction might occur at higher onset potentials in the experimental setting, and C2, with the lowest overpotential among all structures considered in this study, and therefore lying on the top of the volcano.  



\section{Conclusions}
This research has demonstrated that cerium (Ce) doped graphene-based materials can provide a promising avenue for catalyst design. The study's theoretical findings using DFT calculations reveal that Ce doping shows improved catalytic activity towards the ORR, enhancing the potential for clean electricity generation and hydrogen peroxide electrogeneration. Our results are in agreement with previous experimental works in the literature, where the functionalization with cerium and oxygen resulted in carbon-based materials with better performance for the ORR in comparison with their non-functionalized counterparts. Moreover, when considering different structural models of Ce-doped carbon, it was found that certain combinations are more effective in promoting the desired reactions, such as Ce doping when combined with an oxygen atom and located near a border has the potential to be selective for the 2-electron pathway. These results shed light on Ce-doped carbon catalysts and show that understanding the catalytic activity at an atomic level is paramount for the development of novel materials.

\section*{Acknowledgements}

The authors would like to thank FAPESP (2019/01925-4, 2017/10118-0), CNPq (310045/2019-3, 429727/2018-6, 406447/2022-5, 150840/2023-3), CAPES, and the INCT Materials Informatics for the financial support. The authors would also like to thank CCM-UFABC for the computational resources and Tiago Duarte Pereira for the graphical abstract illustration.

\appendix



\bibliographystyle{elsarticle-harv} 
\bibliography{example}

\begin{thebibliography}{30}
\expandafter\ifx\csname natexlab\endcsname\relax\def\natexlab#1{#1}\fi
\providecommand{\url}[1]{\texttt{#1}}
\providecommand{\href}[2]{#2}
\providecommand{\path}[1]{#1}
\providecommand{\DOIprefix}{doi:}
\providecommand{\ArXivprefix}{arXiv:}
\providecommand{\URLprefix}{URL: }
\providecommand{\Pubmedprefix}{pmid:}
\providecommand{\doi}[1]{\href{http://dx.doi.org/#1}{\path{#1}}}
\providecommand{\Pubmed}[1]{\href{pmid:#1}{\path{#1}}}
\providecommand{\bibinfo}[2]{#2}
\ifx\xfnm\relax \def\xfnm[#1]{\unskip,\space#1}\fi
\bibitem[{Assumpção et~al.(2012)Assumpção, Moraes, {De Souza}, Gaubeur, Oliveira, Antonin, Malpass, Rocha, Calegaro, Lanza and Santos}]{cec1}
\bibinfo{author}{Assumpção, M.}, \bibinfo{author}{Moraes, A.}, \bibinfo{author}{{De Souza}, R.}, \bibinfo{author}{Gaubeur, I.}, \bibinfo{author}{Oliveira, R.}, \bibinfo{author}{Antonin, V.}, \bibinfo{author}{Malpass, G.}, \bibinfo{author}{Rocha, R.}, \bibinfo{author}{Calegaro, M.}, \bibinfo{author}{Lanza, M.}, \bibinfo{author}{Santos, M.}, \bibinfo{year}{2012}.
\newblock \bibinfo{title}{Low content cerium oxide nanoparticles on carbon for hydrogen peroxide electrosynthesis}.
\newblock \bibinfo{journal}{Applied Catalysis A: General} \bibinfo{volume}{411-412}, \bibinfo{pages}{1--6}.
\newblock \DOIprefix\doi{https://doi.org/10.1016/j.apcata.2011.09.030}.
\bibitem[{Binnemans and Jones(2015)}]{binnemans2015rare}
\bibinfo{author}{Binnemans, K.}, \bibinfo{author}{Jones, P.T.}, \bibinfo{year}{2015}.
\newblock \bibinfo{title}{Rare earths and the balance problem}.
\newblock \bibinfo{journal}{Journal of Sustainable Metallurgy} \bibinfo{volume}{1}, \bibinfo{pages}{29--38}.
\bibitem[{Chu et~al.(2017)Chu, Cui and Liu}]{sustainable}
\bibinfo{author}{Chu, S.}, \bibinfo{author}{Cui, Y.}, \bibinfo{author}{Liu, N.}, \bibinfo{year}{2017}.
\newblock \bibinfo{title}{The path towards sustainable energy}.
\newblock \bibinfo{journal}{Nature Materials} \bibinfo{volume}{16}, \bibinfo{pages}{16--22}.
\newblock \DOIprefix\doi{https://doi.org/10.1038/nmat4834}.
\bibitem[{Fan et~al.(2023)Fan, Duan, Liu, Mehmood, Qu, Cao, Guo, Zhong and Zhang}]{mcatalysts2}
\bibinfo{author}{Fan, W.}, \bibinfo{author}{Duan, Z.}, \bibinfo{author}{Liu, W.}, \bibinfo{author}{Mehmood, R.}, \bibinfo{author}{Qu, J.}, \bibinfo{author}{Cao, Y.}, \bibinfo{author}{Guo, X.}, \bibinfo{author}{Zhong, J.}, \bibinfo{author}{Zhang, F.}, \bibinfo{year}{2023}.
\newblock \bibinfo{title}{Rational design of heterogenized molecular phthalocyanine hybrid single-atom electrocatalyst towards two-electron oxygen reduction}.
\newblock \bibinfo{journal}{Nature Communications} \bibinfo{volume}{14}, \bibinfo{pages}{1426}.
\newblock \DOIprefix\doi{https://doi.org/10.1038/s41467-023-37066-y}.
\bibitem[{Giannozzi et~al.(2009)Giannozzi, Baroni, Bonini, Calandra, Car, Cavazzoni, Ceresoli, Chiarotti, Cococcioni, Dabo, Corso, de~Gironcoli, Fabris, Fratesi, Gebauer, Gerstmann, Gougoussis, Kokalj, Lazzeri, Martin-Samos, Marzari, Mauri, Mazzarello, Paolini, Pasquarello, Paulatto, Sbraccia, Scandolo, Sclauzero, Seitsonen, Smogunov, Umari and Wentzcovitch}]{Giannozzi_2009}
\bibinfo{author}{Giannozzi, P.}, \bibinfo{author}{Baroni, S.}, \bibinfo{author}{Bonini, N.}, \bibinfo{author}{Calandra, M.}, \bibinfo{author}{Car, R.}, \bibinfo{author}{Cavazzoni, C.}, \bibinfo{author}{Ceresoli, D.}, \bibinfo{author}{Chiarotti, G.L.}, \bibinfo{author}{Cococcioni, M.}, \bibinfo{author}{Dabo, I.}, \bibinfo{author}{Corso, A.D.}, \bibinfo{author}{de~Gironcoli, S.}, \bibinfo{author}{Fabris, S.}, \bibinfo{author}{Fratesi, G.}, \bibinfo{author}{Gebauer, R.}, \bibinfo{author}{Gerstmann, U.}, \bibinfo{author}{Gougoussis, C.}, \bibinfo{author}{Kokalj, A.}, \bibinfo{author}{Lazzeri, M.}, \bibinfo{author}{Martin-Samos, L.}, \bibinfo{author}{Marzari, N.}, \bibinfo{author}{Mauri, F.}, \bibinfo{author}{Mazzarello, R.}, \bibinfo{author}{Paolini, S.}, \bibinfo{author}{Pasquarello, A.}, \bibinfo{author}{Paulatto, L.}, \bibinfo{author}{Sbraccia, C.}, \bibinfo{author}{Scandolo, S.}, \bibinfo{author}{Sclauzero, G.}, \bibinfo{author}{Seitsonen, A.P.}, \bibinfo{author}{Smogunov, A.}, \bibinfo{author}{Umari, P.},
  \bibinfo{author}{Wentzcovitch, R.M.}, \bibinfo{year}{2009}.
\newblock \bibinfo{title}{Quantum espresso: A modular and open-source software project for quantum simulations of materials}.
\newblock \bibinfo{journal}{Journal of Physics: Condensed Matter} \bibinfo{volume}{21}, \bibinfo{pages}{395502}.
\newblock \DOIprefix\doi{10.1088/0953-8984/21/39/395502}.
\bibitem[{Guo et~al.(2023)Guo, Duan, Wu, Zhang, Wang, Zhang, Luo, Lu, Zhang, Mu, Zhang, Han and Wang}]{bn}
\bibinfo{author}{Guo, J.}, \bibinfo{author}{Duan, Y.}, \bibinfo{author}{Wu, T.}, \bibinfo{author}{Zhang, W.}, \bibinfo{author}{Wang, L.}, \bibinfo{author}{Zhang, Y.}, \bibinfo{author}{Luo, Q.}, \bibinfo{author}{Lu, Q.}, \bibinfo{author}{Zhang, Y.}, \bibinfo{author}{Mu, H.}, \bibinfo{author}{Zhang, H.}, \bibinfo{author}{Han, Q.}, \bibinfo{author}{Wang, D.}, \bibinfo{year}{2023}.
\newblock \bibinfo{title}{Atomically dispersed cerium sites in carbon-doped boron nitride for photodriven co2 reduction: Local polarization and mechanism insight}.
\newblock \bibinfo{journal}{Applied Catalysis B: Environmental} \bibinfo{volume}{324}, \bibinfo{pages}{122235}.
\newblock \DOIprefix\doi{https://doi.org/10.1016/j.apcatb.2022.122235}.
\bibitem[{Hwang et~al.(2018)Hwang, Kim, Ryu, Kim, Lee, Kim, Kang, Park, Lanzara, Chung, Mo, Denlinger, Min and Hwang}]{cerium-graphene-2018}
\bibinfo{author}{Hwang, J.}, \bibinfo{author}{Kim, K.}, \bibinfo{author}{Ryu, H.}, \bibinfo{author}{Kim, J.}, \bibinfo{author}{Lee, J.E.}, \bibinfo{author}{Kim, S.}, \bibinfo{author}{Kang, M.}, \bibinfo{author}{Park, B.G.}, \bibinfo{author}{Lanzara, A.}, \bibinfo{author}{Chung, J.}, \bibinfo{author}{Mo, S.K.}, \bibinfo{author}{Denlinger, J.}, \bibinfo{author}{Min, B.I.}, \bibinfo{author}{Hwang, C.}, \bibinfo{year}{2018}.
\newblock \bibinfo{title}{Emergence of kondo resonance in graphene intercalated with cerium}.
\newblock \bibinfo{journal}{Nano Letters} \bibinfo{volume}{18}, \bibinfo{pages}{3661--3666}.
\newblock \DOIprefix\doi{10.1021/acs.nanolett.8b00784}.
\bibitem[{Jain et~al.(2013)Jain, Ong, Hautier, Chen, Richards, Dacek, Cholia, Gunter, Skinner, Ceder and Persson}]{mp}
\bibinfo{author}{Jain, A.}, \bibinfo{author}{Ong, S.P.}, \bibinfo{author}{Hautier, G.}, \bibinfo{author}{Chen, W.}, \bibinfo{author}{Richards, W.D.}, \bibinfo{author}{Dacek, S.}, \bibinfo{author}{Cholia, S.}, \bibinfo{author}{Gunter, D.}, \bibinfo{author}{Skinner, D.}, \bibinfo{author}{Ceder, G.}, \bibinfo{author}{Persson, K.A.}, \bibinfo{year}{2013}.
\newblock \bibinfo{title}{{Commentary: The Materials Project: A materials genome approach to accelerating materials innovation}}.
\newblock \bibinfo{journal}{APL Materials} \bibinfo{volume}{1}, \bibinfo{pages}{011002}.
\newblock \DOIprefix\doi{10.1063/1.4812323}.
\bibitem[{Jiang et~al.(2019)Jiang, Back, Akey, Xia, Hu, Liang, Schaak, Stavitski, N{\o}rskov, Siahrostami et~al.}]{norskov-tm-sac}
\bibinfo{author}{Jiang, K.}, \bibinfo{author}{Back, S.}, \bibinfo{author}{Akey, A.J.}, \bibinfo{author}{Xia, C.}, \bibinfo{author}{Hu, Y.}, \bibinfo{author}{Liang, W.}, \bibinfo{author}{Schaak, D.}, \bibinfo{author}{Stavitski, E.}, \bibinfo{author}{N{\o}rskov, J.K.}, \bibinfo{author}{Siahrostami, S.}, et~al., \bibinfo{year}{2019}.
\newblock \bibinfo{title}{Highly selective oxygen reduction to hydrogen peroxide on transition metal single atom coordination}.
\newblock \bibinfo{journal}{Nature communications} \bibinfo{volume}{10}, \bibinfo{pages}{3997}.
\bibitem[{Lashanizadegan et~al.(2020)Lashanizadegan, Anafcheh, Mirzazadeh and Gholipoor}]{flake}
\bibinfo{author}{Lashanizadegan, M.}, \bibinfo{author}{Anafcheh, M.}, \bibinfo{author}{Mirzazadeh, H.}, \bibinfo{author}{Gholipoor, P.}, \bibinfo{year}{2020}.
\newblock \bibinfo{title}{Efficient cd/ce nanoparticles supported on reduced graphene oxide for the reduction of 4-nitrophenol and the oxidation of olefins: Experimental and theoretical study}.
\newblock \bibinfo{journal}{Materials Research Bulletin} \bibinfo{volume}{125}, \bibinfo{pages}{110773}.
\newblock \DOIprefix\doi{https://doi.org/10.1016/j.materresbull.2020.110773}.
\bibitem[{Li et~al.(2022)Li, He, Li, Li and Zhao}]{LI2022128048}
\bibinfo{author}{Li, C.}, \bibinfo{author}{He, Y.}, \bibinfo{author}{Li, Z.}, \bibinfo{author}{Li, H.}, \bibinfo{author}{Zhao, Y.}, \bibinfo{year}{2022}.
\newblock \bibinfo{title}{Graphene loaded with corrosion inhibitor cerium (iii) cation for enhancing corrosion resistance of waterborne epoxy coating: Physical barrier and self-healing}.
\newblock \bibinfo{journal}{Colloids and Surfaces A: Physicochemical and Engineering Aspects} \bibinfo{volume}{635}, \bibinfo{pages}{128048}.
\newblock \URLprefix \url{https://www.sciencedirect.com/science/article/pii/S0927775721019178}, \DOIprefix\doi{https://doi.org/10.1016/j.colsurfa.2021.128048}.
\bibitem[{Li et~al.(2021)Li, Qin, Xiao, Liang, Xu, Meng, Sarnello, Fang, Li, Ding, Lyu, Zhu, Pan, Hou, Liu, Lin and Shao}]{CeNgraph}
\bibinfo{author}{Li, J.C.}, \bibinfo{author}{Qin, X.}, \bibinfo{author}{Xiao, F.}, \bibinfo{author}{Liang, C.}, \bibinfo{author}{Xu, M.}, \bibinfo{author}{Meng, Y.}, \bibinfo{author}{Sarnello, E.}, \bibinfo{author}{Fang, L.}, \bibinfo{author}{Li, T.}, \bibinfo{author}{Ding, S.}, \bibinfo{author}{Lyu, Z.}, \bibinfo{author}{Zhu, S.}, \bibinfo{author}{Pan, X.}, \bibinfo{author}{Hou, P.X.}, \bibinfo{author}{Liu, C.}, \bibinfo{author}{Lin, Y.}, \bibinfo{author}{Shao, M.}, \bibinfo{year}{2021}.
\newblock \bibinfo{title}{Highly dispersive cerium atoms on carbon nanowires as oxygen reduction reaction electrocatalysts for zn–air batteries}.
\newblock \bibinfo{journal}{Nano Letters} \bibinfo{volume}{21}, \bibinfo{pages}{4508--4515}.
\newblock \DOIprefix\doi{10.1021/acs.nanolett.1c01493}.
\bibitem[{Lucchetti et~al.(2021)Lucchetti, Almeida, {de Almeida}, Autreto, Honorio and Santos}]{LUCCHETTI2021115429}
\bibinfo{author}{Lucchetti, L.E.}, \bibinfo{author}{Almeida, M.O.}, \bibinfo{author}{{de Almeida}, J.M.}, \bibinfo{author}{Autreto, P.A.}, \bibinfo{author}{Honorio, K.M.}, \bibinfo{author}{Santos, M.C.}, \bibinfo{year}{2021}.
\newblock \bibinfo{title}{Density functional theory studies of oxygen reduction reaction for hydrogen peroxide generation on graphene-based catalysts}.
\newblock \bibinfo{journal}{Journal of Electroanalytical Chemistry} \bibinfo{volume}{895}, \bibinfo{pages}{115429}.
\newblock \DOIprefix\doi{https://doi.org/10.1016/j.jelechem.2021.115429}.
\bibitem[{N{\o}rskov et~al.(2004)N{\o}rskov, Rossmeisl, Logadottir, Lindqvist, Kitchin, Bligaard and Jonsson}]{norskov2004}
\bibinfo{author}{N{\o}rskov, J.K.}, \bibinfo{author}{Rossmeisl, J.}, \bibinfo{author}{Logadottir, A.}, \bibinfo{author}{Lindqvist, L.}, \bibinfo{author}{Kitchin, J.R.}, \bibinfo{author}{Bligaard, T.}, \bibinfo{author}{Jonsson, H.}, \bibinfo{year}{2004}.
\newblock \bibinfo{title}{Origin of the overpotential for oxygen reduction at a fuel-cell cathode}.
\newblock \bibinfo{journal}{The Journal of Physical Chemistry B} \bibinfo{volume}{108}, \bibinfo{pages}{17886--17892}.
\newblock \DOIprefix\doi{https://doi.org/10.1021/jp047349j}.
\bibitem[{Perdew et~al.(1996)Perdew, Burke and Ernzerhof}]{perdew1996generalized}
\bibinfo{author}{Perdew, J.P.}, \bibinfo{author}{Burke, K.}, \bibinfo{author}{Ernzerhof, M.}, \bibinfo{year}{1996}.
\newblock \bibinfo{title}{Generalized gradient approximation made simple}.
\newblock \bibinfo{journal}{Physical review letters} \bibinfo{volume}{77}, \bibinfo{pages}{3865}.
\newblock \DOIprefix\doi{https://doi.org/10.1103/PhysRevLett.77.3865}.
\bibitem[{Pinheiro et~al.(2018)Pinheiro, Paz, Aveiro, Parreira, Souza, Camargo and Santos}]{cec}
\bibinfo{author}{Pinheiro, V.S.}, \bibinfo{author}{Paz, E.C.}, \bibinfo{author}{Aveiro, L.R.}, \bibinfo{author}{Parreira, L.S.}, \bibinfo{author}{Souza, F.M.}, \bibinfo{author}{Camargo, P.H.}, \bibinfo{author}{Santos, M.C.}, \bibinfo{year}{2018}.
\newblock \bibinfo{title}{Ceria high aspect ratio nanostructures supported on carbon for hydrogen peroxide electrogeneration}.
\newblock \bibinfo{journal}{Electrochimica Acta} \bibinfo{volume}{259}, \bibinfo{pages}{865--872}.
\newblock \DOIprefix\doi{https://doi.org/10.1016/j.electacta.2017.11.010}.
\bibitem[{Prandini et~al.(2018)Prandini, Marrazzo, Castelli, Mounet and Marzari}]{sssp}
\bibinfo{author}{Prandini, G.}, \bibinfo{author}{Marrazzo, A.}, \bibinfo{author}{Castelli, I.E.}, \bibinfo{author}{Mounet, N.}, \bibinfo{author}{Marzari, N.}, \bibinfo{year}{2018}.
\newblock \bibinfo{title}{Precision and efficiency in solid-state pseudopotential calculations}.
\newblock \bibinfo{journal}{npj Computational Materials} \bibinfo{volume}{4}.
\newblock \DOIprefix\doi{https://doi.org/10.1038/s41524-018-0127-2}.
\bibitem[{Santos et~al.(2022)Santos, Antonin, Souza, Aveiro, Pinheiro, Gentil, Lima, Moura, Silva, Lucchetti, Codognoto, Robles and Lanza}]{fenton}
\bibinfo{author}{Santos, M.C.}, \bibinfo{author}{Antonin, V.S.}, \bibinfo{author}{Souza, F.M.}, \bibinfo{author}{Aveiro, L.R.}, \bibinfo{author}{Pinheiro, V.S.}, \bibinfo{author}{Gentil, T.C.}, \bibinfo{author}{Lima, T.S.}, \bibinfo{author}{Moura, J.P.}, \bibinfo{author}{Silva, C.R.}, \bibinfo{author}{Lucchetti, L.E.}, \bibinfo{author}{Codognoto, L.}, \bibinfo{author}{Robles, I.}, \bibinfo{author}{Lanza, M.R.}, \bibinfo{year}{2022}.
\newblock \bibinfo{title}{Decontamination of wastewater containing contaminants of emerging concern by electrooxidation and fenton-based processes – a review on the relevance of materials and methods}.
\newblock \bibinfo{journal}{Chemosphere} \bibinfo{volume}{307}, \bibinfo{pages}{135763}.
\newblock \DOIprefix\doi{https://doi.org/10.1016/j.chemosphere.2022.135763}.
\bibitem[{Siahrostami et~al.(2013)Siahrostami, Verdaguer-Casadevall, Karamad, Deiana, Malacrida, Wickman, Escudero-Escribano, Paoli, Frydendal, Hansen, Chorkendorff, Stephens and Rossmeisl}]{h2o2}
\bibinfo{author}{Siahrostami, S.}, \bibinfo{author}{Verdaguer-Casadevall, A.}, \bibinfo{author}{Karamad, M.}, \bibinfo{author}{Deiana, D.}, \bibinfo{author}{Malacrida, P.}, \bibinfo{author}{Wickman, B.}, \bibinfo{author}{Escudero-Escribano, M.}, \bibinfo{author}{Paoli, E.A.}, \bibinfo{author}{Frydendal, R.}, \bibinfo{author}{Hansen, T.W.}, \bibinfo{author}{Chorkendorff, I.}, \bibinfo{author}{Stephens, I.E.L.}, \bibinfo{author}{Rossmeisl, J.}, \bibinfo{year}{2013}.
\newblock \bibinfo{title}{Enabling direct h2o2 production through rational electrocatalyst design}.
\newblock \bibinfo{journal}{Nature Materials} \bibinfo{volume}{12}, \bibinfo{pages}{1137--1143}.
\newblock \DOIprefix\doi{https://doi.org/10.1038/nmat3795}.
\bibitem[{Sims et~al.(2022)Sims, Kesler, Henderson, Castillo, Fishman, Weiss, Singleton, Eggert, McCall and Rios}]{cerium}
\bibinfo{author}{Sims, Z.C.}, \bibinfo{author}{Kesler, M.S.}, \bibinfo{author}{Henderson, H.B.}, \bibinfo{author}{Castillo, E.}, \bibinfo{author}{Fishman, T.}, \bibinfo{author}{Weiss, D.}, \bibinfo{author}{Singleton, P.}, \bibinfo{author}{Eggert, R.}, \bibinfo{author}{McCall, S.K.}, \bibinfo{author}{Rios, O.}, \bibinfo{year}{2022}.
\newblock \bibinfo{title}{How cerium and lanthanum as coproducts promote stable rare earth production and new alloys}.
\newblock \bibinfo{journal}{Journal of Sustainable Metallurgy} \bibinfo{volume}{8}, \bibinfo{pages}{1225--1234}.
\bibitem[{Souza et~al.(2022)Souza, Pinheiro, Gentil, Lucchetti, Silva, {L.M.G. Santos}, {De Oliveira}, Dourado, Amaral-Labat, Okamoto and Santos}]{fuelcell}
\bibinfo{author}{Souza, F.M.}, \bibinfo{author}{Pinheiro, V.S.}, \bibinfo{author}{Gentil, T.C.}, \bibinfo{author}{Lucchetti, L.E.}, \bibinfo{author}{Silva, J.}, \bibinfo{author}{{L.M.G. Santos}, M.}, \bibinfo{author}{{De Oliveira}, I.}, \bibinfo{author}{Dourado, W.M.}, \bibinfo{author}{Amaral-Labat, G.}, \bibinfo{author}{Okamoto, S.}, \bibinfo{author}{Santos, M.C.}, \bibinfo{year}{2022}.
\newblock \bibinfo{title}{Alkaline direct liquid fuel cells: Advances, challenges and perspectives}.
\newblock \bibinfo{journal}{Journal of Electroanalytical Chemistry} \bibinfo{volume}{922}, \bibinfo{pages}{116712}.
\newblock \DOIprefix\doi{https://doi.org/10.1016/j.jelechem.2022.116712}.
\bibitem[{Sun et~al.(2023)Sun, Chen, Yu, Yin and Tian}]{cn}
\bibinfo{author}{Sun, D.}, \bibinfo{author}{Chen, Y.}, \bibinfo{author}{Yu, X.}, \bibinfo{author}{Yin, Y.}, \bibinfo{author}{Tian, G.}, \bibinfo{year}{2023}.
\newblock \bibinfo{title}{Engineering high-coordinated cerium single-atom sites on carbon nitride nanosheets for efficient photocatalytic amine oxidation and water splitting into hydrogen}.
\newblock \bibinfo{journal}{Chemical Engineering Journal} \bibinfo{volume}{462}, \bibinfo{pages}{142084}.
\newblock \DOIprefix\doi{https://doi.org/10.1016/j.cej.2023.142084}.
\bibitem[{Tian et~al.(2020)Tian, Lu, Xia and Lou}]{energy}
\bibinfo{author}{Tian, X.}, \bibinfo{author}{Lu, X.F.}, \bibinfo{author}{Xia, B.Y.}, \bibinfo{author}{Lou, X.W.D.}, \bibinfo{year}{2020}.
\newblock \bibinfo{title}{Advanced electrocatalysts for the oxygen reduction reaction in energy conversion technologies}.
\newblock \bibinfo{journal}{Joule} \bibinfo{volume}{4}, \bibinfo{pages}{45--68}.
\newblock \DOIprefix\doi{https://doi.org/10.1016/j.joule.2019.12.014}.
\bibitem[{Wang et~al.(2020)Wang, Zhou, Lin, Yang, Hu and Xie}]{feNOH}
\bibinfo{author}{Wang, F.}, \bibinfo{author}{Zhou, Y.}, \bibinfo{author}{Lin, S.}, \bibinfo{author}{Yang, L.}, \bibinfo{author}{Hu, Z.}, \bibinfo{author}{Xie, D.}, \bibinfo{year}{2020}.
\newblock \bibinfo{title}{Axial ligand effect on the stability of fe–n–c electrocatalysts for acidic oxygen reduction reaction}.
\newblock \bibinfo{journal}{Nano Energy} \bibinfo{volume}{78}, \bibinfo{pages}{105128}.
\newblock \DOIprefix\doi{https://doi.org/10.1016/j.nanoen.2020.105128}.
\bibitem[{Yang et~al.(2023)Yang, Qiao, Li, Wang, Li, Wu and Liu}]{cn2}
\bibinfo{author}{Yang, J.}, \bibinfo{author}{Qiao, F.}, \bibinfo{author}{Li, B.}, \bibinfo{author}{Wang, T.}, \bibinfo{author}{Li, C.}, \bibinfo{author}{Wu, J.}, \bibinfo{author}{Liu, D.}, \bibinfo{year}{2023}.
\newblock \bibinfo{title}{Activating a ceria-doped carbon nitride for elemental mercury adsorption by constructing a heterojunction interface: A dft guided experimental study}.
\newblock \bibinfo{journal}{Separation and Purification Technology} \bibinfo{volume}{324}, \bibinfo{pages}{124649}.
\newblock \DOIprefix\doi{https://doi.org/10.1016/j.seppur.2023.124649}.
\bibitem[{Zacherle et~al.(2013)Zacherle, Schriever, De~Souza and Martin}]{hub}
\bibinfo{author}{Zacherle, T.}, \bibinfo{author}{Schriever, A.}, \bibinfo{author}{De~Souza, R.A.}, \bibinfo{author}{Martin, M.}, \bibinfo{year}{2013}.
\newblock \bibinfo{title}{Ab initio analysis of the defect structure of ceria}.
\newblock \bibinfo{journal}{Phys. Rev. B} \bibinfo{volume}{87}, \bibinfo{pages}{134104}.
\newblock \DOIprefix\doi{10.1103/PhysRevB.87.134104}.
\bibitem[{Zhang et~al.(2017)Zhang, Sha, Fei, Liu, Yazdi, Zhang, Zhong, Zou, Zhao, Yu, Jiang, Ringe, Yakobson, Dong, Chen and Tour}]{ruNOH}
\bibinfo{author}{Zhang, C.}, \bibinfo{author}{Sha, J.}, \bibinfo{author}{Fei, H.}, \bibinfo{author}{Liu, M.}, \bibinfo{author}{Yazdi, S.}, \bibinfo{author}{Zhang, J.}, \bibinfo{author}{Zhong, Q.}, \bibinfo{author}{Zou, X.}, \bibinfo{author}{Zhao, N.}, \bibinfo{author}{Yu, H.}, \bibinfo{author}{Jiang, Z.}, \bibinfo{author}{Ringe, E.}, \bibinfo{author}{Yakobson, B.I.}, \bibinfo{author}{Dong, J.}, \bibinfo{author}{Chen, D.}, \bibinfo{author}{Tour, J.M.}, \bibinfo{year}{2017}.
\newblock \bibinfo{title}{Single-atomic ruthenium catalytic sites on nitrogen-doped graphene for oxygen reduction reaction in acidic medium}.
\newblock \bibinfo{journal}{ACS Nano} \bibinfo{volume}{11}, \bibinfo{pages}{6930--6941}.
\newblock \DOIprefix\doi{10.1021/acsnano.7b02148}.
\bibitem[{Zhang et~al.(2023)Zhang, Liu, Li, Jian, Gao, Lu and Yue}]{mcatalysts}
\bibinfo{author}{Zhang, X.}, \bibinfo{author}{Liu, J.}, \bibinfo{author}{Li, R.}, \bibinfo{author}{Jian, X.}, \bibinfo{author}{Gao, X.}, \bibinfo{author}{Lu, Z.}, \bibinfo{author}{Yue, X.}, \bibinfo{year}{2023}.
\newblock \bibinfo{title}{Machine learning screening of high-performance single-atom electrocatalysts for two-electron oxygen reduction reaction}.
\newblock \bibinfo{journal}{Journal of Colloid and Interface Science} \bibinfo{volume}{645}, \bibinfo{pages}{956--963}.
\newblock \DOIprefix\doi{https://doi.org/10.1016/j.jcis.2023.05.011}.
\bibitem[{Zhang et~al.(2011)Zhang, Zheng and Jiang}]{nanotube}
\bibinfo{author}{Zhang, Z.W.}, \bibinfo{author}{Zheng, W.T.}, \bibinfo{author}{Jiang, Q.}, \bibinfo{year}{2011}.
\newblock \bibinfo{title}{Hydrogen adsorption on ce/swcnt systems: a dft study}.
\newblock \bibinfo{journal}{Phys. Chem. Chem. Phys.} \bibinfo{volume}{13}, \bibinfo{pages}{9483--9489}.
\newblock \DOIprefix\doi{https://doi.org/10.1039/C0CP02917C}.
\bibitem[{Zhao et~al.(2018)Zhao, Dong, Han, Zhao and Tang}]{Cu-Ce-nano}
\bibinfo{author}{Zhao, Y.}, \bibinfo{author}{Dong, F.}, \bibinfo{author}{Han, W.}, \bibinfo{author}{Zhao, H.}, \bibinfo{author}{Tang, Z.}, \bibinfo{year}{2018}.
\newblock \bibinfo{title}{Construction of cu--ce/graphene catalysts via a one-step hydrothermal method and their excellent co catalytic oxidation performance}.
\newblock \bibinfo{journal}{RSC advances} \bibinfo{volume}{8}, \bibinfo{pages}{1583--1592}.

\end{thebibliography}






\end{document}